\documentclass[11pt]{article}

\usepackage[a4paper,margin=1in]{geometry}

\usepackage[T1]{fontenc}

\usepackage[utf8]{inputenc}

\usepackage{mathptmx}

\usepackage{microtype}

\usepackage{graphicx}

\usepackage{booktabs}

\usepackage{array}

\usepackage{adjustbox}

\usepackage{caption}
\usepackage{setspace}
\usepackage[round,authoryear]{natbib}
\setcitestyle{aysep={,},yysep={;}}
\usepackage{float}

\usepackage{url}

\usepackage[hidelinks]{hyperref}

\usepackage{authblk}

\graphicspath{{figures/}}
\title{Comparative Validation of GPT-4o-mini and Teacher Mean Scores for Automated Scoring of Music Analysis Responses: Single-Pass Deployment, Repeatability, and Strategy-Specific Bias}

\author[1]{Baicheng Lin}

\author[2]{Lingxi Jin}

\author[1]{Kyung-Seok Min}

\affil[1]{Department of Education, Sejong University, South Korea}

\affil[2]{Department of Educational Technology, Ewha Womans University, South Korea}

\date{}

\begin{document}

\maketitle

\begin{abstract}

Scoring open-ended music analysis responses is time-consuming and requires nuanced judgments of harmonic knowledge and formal understanding. This study evaluates the validity and repeatability of GPT-4o-mini for rubric-based scoring of music analysis essays, benchmarking against teacher mean scores. A dataset of 300 university-level student responses was scored by teachers on four dimensions, Harmony, Form, Reasoning, and Terminology. Each dimension was rated on a five-point scale, yielding total scores ranging from 4 to 20. GPT-4o-mini scored responses using three prompting strategies: few-shot prompting combined with chain-of-thought reasoning (Fs+CoT), retrieval-augmented generation (RAG), and self-consistency based on five internal generations per administration (SC, m = 5). Each strategy was administered three times with the model, prompt, rubric, and response held constant. Single-Pass Deployment (Run1) scores represented an operational condition, whereas the median of the median score across three independent runs (Median3R) was used to examine aggregation robustness. Agreement with teacher mean scores were evaluated using Pearson's r, ICC (2,1), Krippendorff's $\alpha$, quadratic weighted kappa, and RMSE. Cross-run repeatability within the three-run condition (3R) condition assessed using intra-LLM ICCs and within-response variability. In single-administration conditions, Fs+CoT showed the strongest agreement with teacher mean scores (ICC (2,1) = 0.657, QWK = 0.656), whereas RAG produced the largest RMSE (2.817). SC showed weaker agreement with teacher mean scores (ICC (2,1) = 0.537), despite stable scores. Median3R aggregation produced only minor changes and did not alter the strategy ordering. Fs+CoT remained the most strongly aligned strategy (ICC (2,1) = 0.656, QWK = 0.655). All strategies demonstrated cross three-runs repeatability, with only minor differences in intra-LLM ICCs across the three prompting strategies. SC was the most reproducible, although its individual-level agreement with teacher scores remained limited. Systematic bias differed across strategies. Fs+CoT under-scored relative to teacher mean scores (bias = -1.730), RAG over-scored (bias = 1.787), and SC showed minimal directional bias (bias = -0.083). Form and Reasoning tended to receive lower scores than Harmony. GPT-4o-mini can generate highly repeatable scores for complex music analysis responses, but prompting strategies produce distinct scoring profiles. Fs+CoT offers the strongest agreement with teacher mean scores and SC maximizes repeatability and minimizes directional bias. Operational use therefore requires strategy-specific calibration, dimension-level validation, and continued human oversight.

\end{abstract}


\section{Background and Purpose}

Assessing students' written analyses of musical works is an important component of music theory education \citep{cook1994guide}. These assessments require students to identify harmonic functions, describe formal structures, apply appropriate terminology, and justify their interpretations through coherent theoretical reasoning \citep{kostka2018tonal}. Unlike selected response tests, music analysis essays reveal not only whether students know particular concepts, but also how they organize evidence, connect musical features, and construct discipline-specific arguments. They therefore provide valuable information about students' conceptual understanding, analytical reasoning, and disciplinary literacy.

Despite their educational value, open-ended music analysis tasks are difficult to score efficiently and consistently. Teachers must interpret responses that may differ substantially in wording, structure, theoretical depth, and analytical approach. Differences in how raters interpret and apply the rubric can reduce scoring consistency. Previous research has shown that performance-based and open-response assessments are susceptible to inter-rater variability, inconsistent application of scoring criteria, and cognitive judgment biases \citep{vanderschaaf2012criteria}. These problems become more pronounced in large courses, where teachers must evaluate many responses within limited time and provide individualized feedback. Establishing a fair, reliable, and scalable scoring process remains a major challenge in educational assessment.

Recent developments in large language models (LLMs) have introduced new possibilities for automating the evaluation of complex, open-response tasks. Automated scoring systems, online assessment platforms, and AI-supported feedback tools have the potential to reduce teachers' grading workload while improving the consistency with which scoring criteria are applied \citep{dikli2006overview,jong2023feedback}. In these fields, computational systems have been used to classify responses, identify relevant content, and estimate scores based on predefined rubrics. However, the use of automated scoring in music education remains limited, particularly for written music analysis tasks that require the integration of theoretical terminology, structural interpretation, and explanatory reasoning \citep{lin2025llms}.

Music analysis presents specific challenges for automated assessment. A written analysis may refer to harmony, form, thematic relationships, tonal direction, and expressive function at the same time. In educational settings, students' written analyses provide insight into their conceptual understanding, reasoning strategies, and disciplinary literacy \citep{hennessy2021chasing}. A response may also be theoretically plausible even when its wording differs from a model answer. An automated scoring system must evaluate more than surface-level keyword overlap. It must determine whether the student has used musical concepts accurately, described structural relationships coherently, and supported analytical claims with relevant evidence \citep{lin2025llms}. These characteristics make music analysis essays a suitable but demanding context for examining the validity of large language model--based scoring.

Prompt design and score stability are central methodological considerations when LLMs are used as automated evaluators. Few-shot examples help the model interpret how a scoring rubric should be applied, while Chain-of-Thought prompting supports criterion-based reasoning before a final score is assigned \citep{brown2020language,wei2022chain}. Retrieval-Augmented Generation provides domain-specific reference materials that may strengthen the model's use of disciplinary knowledge and scoring criteria \citep{lewis2020retrieval}. Self-Consistency, in contrast, generates multiple sampled judgments and aggregates them to reduce dependence on a single stochastic output \citep{wang2022selfconsistency}. Nevertheless, sampling-based generation may produce different scores across repeated runs even when the model, prompt, rubric, and student response remain unchanged. A single model call therefore represents an efficient and realistic deployment condition, whereas repeated scoring and aggregation provide a more robust basis for examining score stability and reproducibility \citep{schroeder2024trust,garciavarela2025chatgpt}.

The present study examines the use of GPT-4o-mini for scoring students' written analyses of musical works. Teacher mean scores are used as an external reference because they represent the combined judgment of human evaluators and reduce dependence on a single teacher's rating. The study systematically compares three prompting strategies: Few-shot plus Chain-of-Thought prompting (Fs+CoT), Retrieval-Augmented Generation (RAG), and Self-Consistency with five generated evaluations (SC, m = 5). The three strategies were compared with the teacher mean scores in terms of validity, agreement, reliability, scoring accuracy, and systematic bias.

The study also distinguishes between two scoring scenarios. The first is the first model run, referred to as Run1, in which each response received one model-generated score. The second is a three independent run, referred to as three-run condition (3R). Median and Mean score across three independent runs referred to as Median3R and Mean3R. Across these runs, the model, prompt, rubric, and input response remain constant, while only sampling stochasticity varies at a temperature of 0.9 \citep{renze2024temperature}. This design makes it possible to examine whether repeated scoring and median aggregation improve the stability and accuracy of LLM-based evaluation without altering the underlying scoring procedure.

By comparing prompting strategies and reporting both Run1 and 3R results, this study addresses two related questions: how closely GPT-4o-mini can reproduce teacher-based evaluations of music analysis essays, and how stable its scoring performance remains across repeated generations. They contribute to the development of more transparent, reproducible, and data-informed assessment practices in music theory education.

\section{Research Questions}

To achieve these goals, we address the following research questions:

\begin{description}
\item[RQ1a.] Under Run1, to what extent do the three prompting strategies (Fs+CoT, RAG, and SC) agree with teacher mean scores at both the dimension and total-score levels?
\item[RQ1b.] When scores from three independent runs are aggregated using the median, does agreement with the teacher mean scores improve, and is the relative ordering of the three prompting strategies preserved?
\item[RQ2.] With the prompts held constant and only sampling stochasticity varied, how repeatable are the scores within each prompting strategy across 3R, and how does run-to-run variability differ across responses?
\item[RQ3.] To what extent do the three prompting strategies exhibit systematic severity or leniency relative to the teacher mean scores, and do these scoring biases vary across the four rubric dimensions?
\end{description}

\section{Method}

\subsection{Sample and Assessment Task}

This study was conducted at a normal university in China within the School of Music. A total of 300 written responses to a music analysis task were collected from undergraduate students enrolled in music-related programs through an online assessment platform. The task required students to analyze Piano Sonata No. 6 in D major, K. 284, movement 3a, focusing on harmonic progressions and formal structures.

\subsection{Teacher Scoring Procedure}

Three experienced music theory teachers independently rated all responses using a common analytic rubric. The rubric consisted of four dimensions: (1) accuracy of harmonic analysis, (2) identification of musical form and structural organization, (3) logical coherence and completeness of reasoning, and (4) precision in the use of music-theoretical terminology. Each dimension was rated on a five-point scale (total score =4-20), and the mean of the three teacher ratings was used as the reference score for evaluating the AI-generated scores.

\subsection{LLM Scoring Conditions and Prompting Strategies}

GPT-4o-mini scored each response under three prompting strategies: few-shot prompting combined with chain-of-thought reasoning (Fs+CoT), retrieval-augmented generation (RAG), and self-consistency based on five internal generations per scoring administration (SC, m = 5). Each strategy was independently administered three times for every response. Across runs, the model, prompt, rubric, and student response were held constant, while sampling stochasticity was maintained at a temperature of 0.9.

\subsection{Single-Run and Score Aggregation}

For operational validity evidence, we additionally report agreement based on Run1. Run 1 represented a single-administration condition in which each response was scored once by each prompting strategy. For robust reporting, the primary AI score was Median3R, was used as the primary aggregated score. Median aggregation was adopted as a study-specific robustness procedure to reduce the influence of an unusually high or low score from a single stochastic run. Although this procedure was informed by research on repeated sampling and self-consistency in LLM outputs, it was not treated as an established standard for automated scoring \citep{gaggioli2025assessing}.

\subsection{Association, Agreement, and Scoring Error}

Association between AI and teacher mean scores was evaluated using Pearson's r. Absolute agreement was assessed using the intraclass correlation coefficient, ICC (2, 1), based on a two-way random-effects, absolute-agreement model, as well as Krippendorff's $\alpha$ for interval-level data and quadratic weighted kappa (QWK). Scoring error was evaluated using root mean square error (RMSE) and mean absolute error (MAE). Mean signed bias was also calculated as the AI score minus the corresponding teacher mean score.

\subsection{Repeatability and Bias Analysis}

Within each strategy, repeatability across the 3R was estimated using intra-LLM ICC (2,1) and ICC (2,3), and the per-response SD across runs was summarized to characterize stability.

To examine scoring bias and dimension-specific discrepancies, signed difference scores were calculated by subtracting the teacher mean scores from the corresponding AI score for each response and rubric dimension. The mean signed difference was interpreted as scoring bias, with negative values indicating relative under-scoring or greater severity and positive values indicating relative over-scoring or greater leniency \citep{ramineni2013automated}. For each prompting strategy and rubric dimension, mean signed bias, mean absolute error (MAE), root mean square error (RMSE), exact agreement, and the proportions of under-scoring and over-scoring were calculated.

Statistical uncertainty was evaluated using 95\% bootstrap confidence intervals obtained by resampling student responses with replacement. Differences between prompting strategies were assessed using paired bootstrap confidence intervals for metric differences. Differences between prompting paradigms were estimated using paired bootstrap confidence intervals based on the same student responses, thereby preserving the paired structure of the data \citep{bestgen2022difference}.

\section{Results}

\subsection{RQ1a: Agreement with Teacher Scores in Run1}

As shown in Table 1, of the three prompting strategies under Run1. Fs+CoT achieved the strongest agreement with teacher mean scores (r = 0.795, ICC (2,1) = 0.657), followed by RAG (r=0.669) and Self-Consistency (SC) (r = 0.545). Fs+CoT also obtained the highest Krippendorff's $\alpha$ (0.625) and QWK (0.656) with the lowest RMSE (2.461), indicating stronger agreement with the teacher mean scores and lower score-level error under the Run1. These results indicate that Fs+CoT most closely approximated the teacher mean scores among the three strategies. RAG and SC display moderate and lower consistency, respectively.

\begin{table}[htbp]
\centering
\caption{Inter-Rater Reliability and Agreement Across Prompt Strategies}
\resizebox{\textwidth}{!}{%
\begin{tabular}{lccccc}
\toprule
Strategies & Pearson r & ICC(2,1) & Krippendorff's $\alpha$ & QWK & RMSE \\
\midrule
Fs+CoT & 0.795 & 0.657 & 0.625 & 0.656 & 2.461 \\
RAG & 0.669 & 0.525 & 0.468 & 0.525 & 2.817 \\
SC & 0.545 & 0.537 & 0.536 & 0.536 & 2.718 \\
\bottomrule
\end{tabular}%
}
\end{table}

At the dimension level (Table 2), Fs+CoT showed the strongest overall alignment with teacher mean scores. Agreement was highest for Form and Reasoning, with ICC (2,1) and QWK values of approximately 0.80. These ICC values indicate good reliability according to commonly used interpretive guidelines \citep{koo2016guideline}. RMSE was also lowest for Form (0.918) and Reasoning (0.943), indicating smaller score-level errors in these dimensions. Agreement was weaker for Harmony and Terminology. Fs+CoT showed negative bias across all four dimensions, indicating a consistent tendency to assign lower scores than the teacher mean scores.

RAG demonstrated its strongest dimension-level performance in Reasoning, with ICC (2,1) =0.716, QWK=0.716, and RMSE=1.122. Agreement was lower for Form, Harmony, and Terminology. Positive bias was observed in all four dimensions indicating systematic score inflation relative to the teacher raters. Such mean differences are important because automated and human scores may show moderate agreement while remaining misaligned in score level \citep{williamson2012framework}.

SC also performed best in Reasoning, with ICC (2,1) =0.687, QWK=0.687, and RMSE=1.190. Its agreement was lower for Form and Harmony and weakest for Terminology. Although SC showed minimal directional bias across dimensions, with values ranging from -0.103 to 0.073, its RMSE values remained comparatively high. Near-zero means bias therefore did not imply close agreement at the individual-response level. Positive and negative scoring errors may have offset one another when averaged, while substantial response-level discrepancies remained.

\begin{table}[htbp]
\centering
\caption{Dimension-wise agreement between prompt strategies and teacher mean scores}
\resizebox{\textwidth}{!}{%
\begin{tabular}{lccccccc}
\toprule
Strategies & Dimension & r & ICC (2,1) & Krippendorff's $\alpha$ & QWK & RMSE & Bias \\
\midrule
Fs+CoT & Harmony & 0.753 & 0.698 & 0.689 & 0.698 & 1.007 & -0.433 \\
Fs+CoT & Form & 0.830 & 0.802 & 0.798 & 0.801 & 0.918 & -0.383 \\
Fs+CoT & Reasoning & 0.827 & 0.802 & 0.798 & 0.801 & 0.943 & -0.377 \\
Fs+CoT & Terminology & 0.738 & 0.661 & 0.645 & 0.660 & 1.044 & -0.530 \\
RAG & Harmony & 0.662 & 0.610 & 0.594 & 0.609 & 1.080 & 0.480 \\
RAG & Form & 0.715 & 0.671 & 0.661 & 0.671 & 1.179 & 0.510 \\
RAG & Reasoning & 0.754 & 0.716 & 0.709 & 0.716 & 1.122 & 0.473 \\
RAG & Terminology & 0.610 & 0.567 & 0.552 & 0.566 & 1.117 & 0.433 \\
SC & Harmony & 0.540 & 0.529 & 0.528 & 0.528 & 1.238 & -0.047 \\
SC & Form & 0.629 & 0.627 & 0.627 & 0.626 & 1.281 & 0.073 \\
SC & Reasoning & 0.687 & 0.687 & 0.687 & 0.687 & 1.190 & -0.010 \\
SC & Terminology & 0.472 & 0.458 & 0.457 & 0.457 & 1.300 & -0.103 \\
\bottomrule
\end{tabular}%
}
\end{table}

Across all three prompting strategies, Reasoning generally showed the strongest agreement with teacher mean scores. Terminology showed the lowest agreement for all three strategies. It also had the highest RMSE for Fs+CoT and SC, whereas Form had the highest RMSE for RAG. This pattern indicates that LLM scoring performance was dimension-specific rather than uniform across the rubric. Previous analytic automated-scoring studies have similarly emphasized that performance should be evaluated separately for each rubric dimension because different dimensions involve distinct linguistic, conceptual, and evidential requirements \citep{rahimi2017assessing,yoo2025dataset}. Accordingly, dimension-level results provide information that cannot be captured by total-score agreement alone.

\subsection{RQ1b: Effects of Median3R Aggregation on Agreement}

As shown in Table 3, aggregating 3R using the Median3R resulted in only modest changes in agreement with teacher mean scores while preserving the performance pattern observed under single-pass deployment. Across the three prompting strategies, Fs+CoT remained the best-performing strategy, followed by RAG and SC. Thus, score aggregation did not alter the relative ordering of the prompting strategies.

At the total-score level, Fs+CoT showed virtually identical results to those obtained from Run1. Pearson's correlation remained at 0.795. RMSE decreased only slightly from 2.461 to 2.457. The similarity between the Run1 and Median3R results indicates that aggregation had little effect on Fs+CoT agreement with the teacher mean scores.

Compared with Run1, RAG showed the largest benefit from score aggregation. Although the magnitude of improvement was modest, the consistent direction across all agreement indices suggests that Median3R aggregation reduced stochastic scoring variation for RAG. Nevertheless, positive bias remained substantial (Bias = 1.787), indicating that aggregation did not eliminate the systematic tendency to assign higher scores than the teacher mean scores.

SC exhibited almost no change after aggregation. QWK increased only marginally (0.536 to 0.538), whereas RMSE decreased slightly (2.718 to 2.711). These minimal differences suggest that repeated sampling had little influence on the final scores.

At the dimension level, aggregation produced only small changes, and the overall pattern of results remained unchanged. The Median3R results were very similar to those obtained from Run 1. Improvements were most apparent for RAG, whereas Fs+CoT and SC changed very little. Importantly, aggregation did not change the relative performance pattern across prompting paradigms or rubric dimensions.

\begin{table}[htbp]
\centering
\caption{Association, Agreement, and Error for Median3R Scores Relative to teacher mean scores}
\resizebox{\textwidth}{!}{%
\begin{tabular}{lccccccc}
\toprule
Strategies & Dimension & r & ICC (2,1) & Krippendorff's $\alpha$ & QWK & RMSE & Bias \\
\midrule
Fs+CoT & Harmony & 0.745 & 0.691 & 0.681 & 0.690 & 1.013 & -0.440 \\
Fs+CoT & Form & 0.832 & 0.804 & 0.800 & 0.804 & 0.913 & -0.380 \\
Fs+CoT & Reasoning & 0.828 & 0.802 & 0.799 & 0.801 & 0.942 & -0.380 \\
Fs+CoT & Terminology & 0.738 & 0.661 & 0.645 & 0.660 & 1.044 & -0.530 \\
Fs+CoT & Total & 0.795 & 0.656 & 0.623 & 0.655 & 2.457 & -1.730 \\
RAG & Harmony & 0.662 & 0.611 & 0.595 & 0.610 & 1.079 & 0.477 \\
RAG & Form & 0.711 & 0.670 & 0.660 & 0.669 & 1.170 & 0.490 \\
RAG & Reasoning & 0.741 & 0.705 & 0.698 & 0.705 & 1.133 & 0.457 \\
RAG & Terminology & 0.614 & 0.573 & 0.560 & 0.572 & 1.103 & 0.417 \\
RAG & Total & 0.676 & 0.541 & 0.489 & 0.540 & 2.704 & 1.787 \\
SC & Harmony & 0.540 & 0.529 & 0.528 & 0.528 & 1.238 & -0.047 \\
SC & Form & 0.629 & 0.627 & 0.627 & 0.626 & 1.281 & 0.073 \\
SC & Reasoning & 0.687 & 0.687 & 0.687 & 0.687 & 1.190 & -0.010 \\
SC & Terminology & 0.472 & 0.458 & 0.457 & 0.457 & 1.300 & -0.103 \\
SC & Total & 0.547 & 0.539 & 0.538 & 0.538 & 2.711 & -0.083 \\
\bottomrule
\end{tabular}%
}
\end{table}

\subsection{RQ2: Score Stability across 3R}

Figures 1 and 2, together with Table 4, show clear strategy-specific scoring patterns and high within-strategy repeatability across the 3R. The run-specific distributions largely overlapped for all three prompting strategies. This visual pattern was supported by high ICCs, mean pairwise QWK values, and low within-response standard deviations.

Fs+CoT showed a consistent shift toward lower scores. At the dimension level, this pattern was most visible in Harmony, Reasoning, and Terminology, where Fs+CoT assigned more scores in the lower categories than the teacher mean scores. The same pattern was observed for total scores. The Fs+CoT distribution was shifted to the left of the teacher distribution, indicating greater scoring severity. The 3R Fs+CoT showed similar peaks and distributional shapes. This suggests that the lower scores reflected a systematic severity effect rather than substantial run-to-run variation.

\begin{figure}[htbp]

\centering

\includegraphics[width=0.95\textwidth]{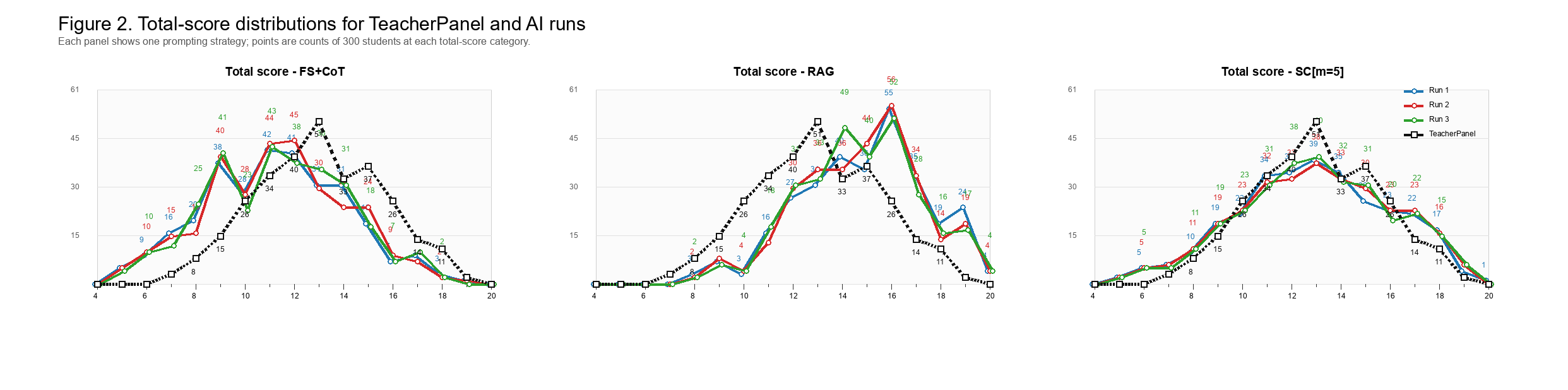}

\caption{Total score distributions for prompt strategies and teacher mean scores}

\end{figure}

\begin{figure}[htbp]

\centering

\includegraphics[width=0.95\textwidth]{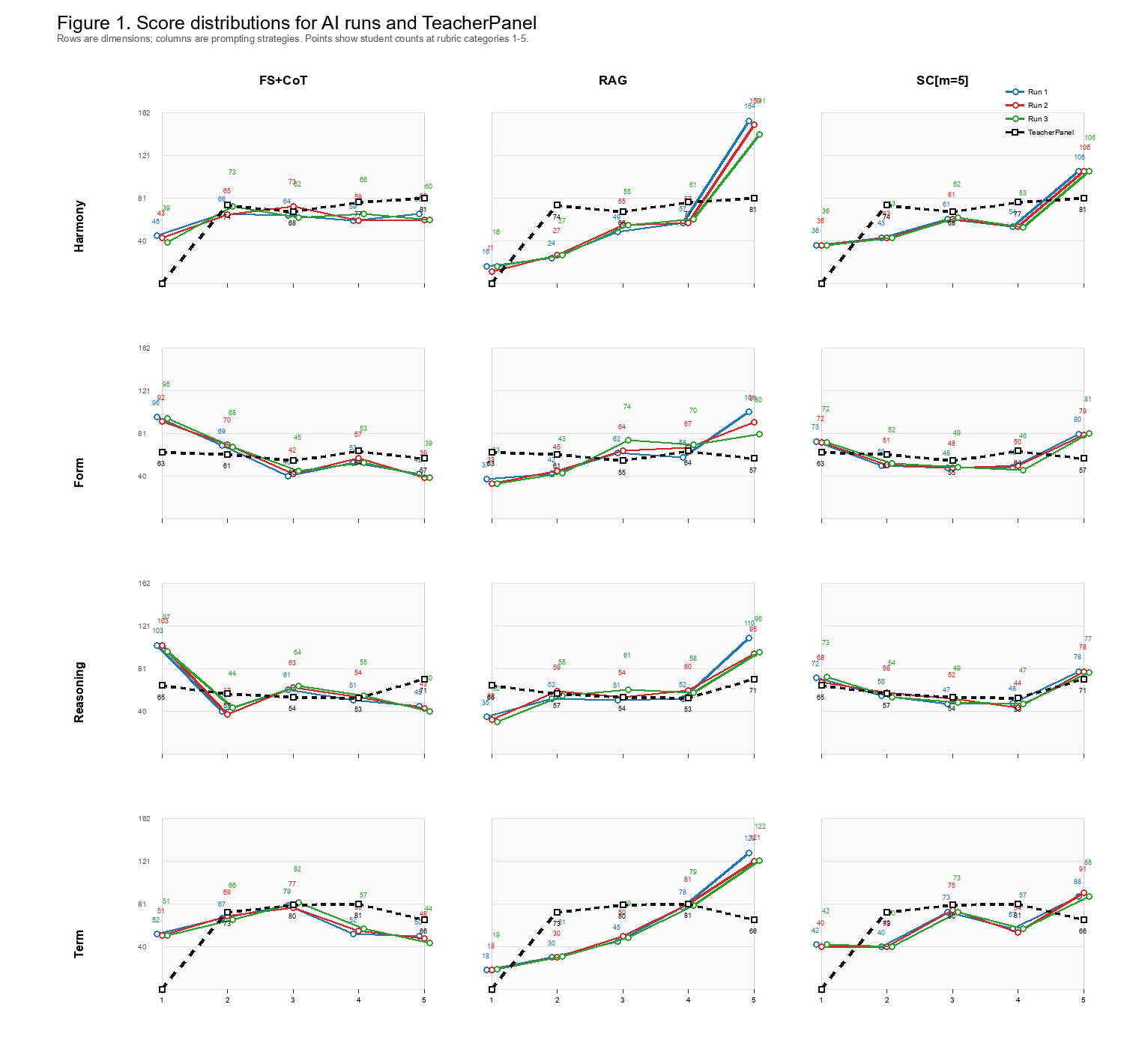}

\caption{Dimension-wise score distribution for prompt strategies and teacher mean scores}

\end{figure}

RAG showed the opposite pattern. Its dimension-level distributions were more concentrated in the upper score categories, particularly in Harmony and Terminology. The total-score distribution was also shifted to the right, with a larger number of responses in the high-score range. Although the 3R RAG was generally similar, some variation was observed at individual score points.

SC score distribution was closest to that of the teacher mean scores. The mean within-response SDs were very small across dimensions (0.010--0.044), and the total-score SD was 0.098. The center of the total-score distribution was also close to that of the teacher scores. These findings show that SC produced highly stable scores across repeated runs. However, similarity in overall distribution does not necessarily imply strong response-level agreement with the teacher mean scores, because score differences may still occur for individual responses.

\begin{table}[htbp]
\centering
\caption{Within-strategy repeatability across 3R}
\resizebox{\textwidth}{!}{%
\begin{tabular}{lcccccc}
\toprule
Strategies & Dimension & ICC (2,1) & ICC (2,3) & Krippendorff's $\alpha$ & Mean pairwise QWK & Mean within-response SD \\
\midrule
Fs+CoT & Harmony & 0.945 & 0.981 & 0.945 & 0.945 & 0.151 \\
Fs+CoT & Form & 0.975 & 0.992 & 0.975 & 0.975 & 0.079 \\
Fs+CoT & Reasoning & 0.971 & 0.990 & 0.971 & 0.971 & 0.083 \\
Fs+CoT & Term & 0.978 & 0.993 & 0.978 & 0.978 & 0.048 \\
Fs+CoT & Total & 0.970 & 0.990 & 0.970 & 0.970 & 0.307 \\
RAG & Harmony & 0.930 & 0.975 & 0.929 & 0.929 & 0.109 \\
RAG & Form & 0.909 & 0.968 & 0.909 & 0.908 & 0.178 \\
RAG & Reasoning & 0.919 & 0.971 & 0.918 & 0.918 & 0.174 \\
RAG & Term & 0.955 & 0.984 & 0.954 & 0.954 & 0.074 \\
RAG & Total & 0.928 & 0.975 & 0.928 & 0.928 & 0.360 \\
SC & Harmony & 0.997 & 0.999 & 0.997 & 0.997 & 0.010 \\
SC & Form & 0.986 & 0.995 & 0.986 & 0.986 & 0.033 \\
SC & Reasoning & 0.985 & 0.995 & 0.985 & 0.985 & 0.044 \\
SC & Term & 0.991 & 0.997 & 0.991 & 0.991 & 0.025 \\
SC & Total & 0.990 & 0.997 & 0.990 & 0.990 & 0.098 \\
\bottomrule
\end{tabular}%
}
\end{table}

The evidence from Figures 1 and 2 and Table 4 indicates that within-strategy repeatability was high for all three prompting strategies. SC showed the strongest run-to-run stability, followed by Fs+CoT and RAG. The differences between AI and teacher scores therefore appear to reflect stable, strategy-specific scoring tendencies rather than random run-to-run variation. Fs+CoT was relatively severe, RAG was relatively lenient, and SC was closest to the teacher mean scores in overall score location. The dimension-level results also show that scoring behavior was not uniform across rubric components.

\subsection{RQ3: Scoring Bias across Strategies and Rubric Dimensions}

RAG showed a consistent positive bias in all four dimensions. All 95\% bootstrap confidence intervals were above zero. This indicates systematic leniency relative to the teacher mean scores. Over-scoring occurred in 44.7\% to 46.0\% of responses across dimensions.

SC showed little directional bias. The bias values ranged from -0.103 to 0.073. The 95\% bootstrap confidence intervals included zero in all four dimensions. This means that SC did not show clear systematic severity or leniency. However, its error indices were relatively large. MAE ranged from 0.817 to 0.963, and RMSE ranged from 1.190 to 1.300.

\begin{table}[htbp]
\centering
\caption{Strategy-level and dimension-specific scoring bias}
\resizebox{\textwidth}{!}{%
\begin{tabular}{lcccccccc}
\toprule
Strategies & Dimension & Bias [95\% bootstrap CI] & Bias [95\% bootstrap CI] & MAE & RMSE & Exact agreement (\%) & Over-scoring (\%) & Under-scoring (\%) \\
\midrule
Fs+CoT & Harmony & -0.440 & [-0.543, -0.337] & 0.733 & 1.013 & 40.0 & 14.0 & 46.0 \\
Fs+CoT & Form & -0.380 & [-0.473, -0.287] & 0.607 & 0.913 & 50.0 & 10.7 & 39.3 \\
Fs+CoT & Reasoning & -0.380 & [-0.480, -0.283] & 0.607 & 0.942 & 50.7 & 11.0 & 38.3 \\
Fs+CoT & Term & -0.530 & [-0.630, -0.423] & 0.790 & 1.044 & 35.0 & 12.7 & 52.3 \\
Fs+CoT & Total & -1.730 & [-1.927, -1.533] & 1.990 & 2.457 & 12.7 & 10.0 & 77.3 \\
RAG & Harmony & 0.477 & [0.367, 0.583] & 0.757 & 1.079 & 42.0 & 45.7 & 12.3 \\
RAG & Form & 0.490 & [0.367, 0.610] & 0.817 & 1.170 & 41.0 & 46.0 & 13.0 \\
RAG & Reasoning & 0.457 & [0.343, 0.577] & 0.790 & 1.133 & 41.0 & 44.7 & 14.3 \\
RAG & Term & 0.417 & [0.297, 0.537] & 0.803 & 1.103 & 38.0 & 45.7 & 16.3 \\
RAG & Total & 1.787 & [1.553, 2.017] & 2.147 & 2.704 & 16.3 & 73.7 & 10.0 \\
SC & Harmony & -0.047 & [-0.190, 0.093] & 0.887 & 1.238 & 37.0 & 28.7 & 34.3 \\
SC & Form & 0.073 & [-0.070, 0.220] & 0.907 & 1.281 & 38.3 & 33.0 & 28.7 \\
SC & Reasoning & -0.010 & [-0.150, 0.123] & 0.817 & 1.190 & 41.7 & 30.7 & 27.7 \\
SC & Term & -0.103 & [-0.247, 0.043] & 0.963 & 1.300 & 33.7 & 32.7 & 33.7 \\
SC & Total & -0.083 & [-0.383, 0.223] & 2.143 & 2.711 & 13.3 & 44.0 & 42.7 \\
\bottomrule
\end{tabular}%
}
\end{table}

Figures 3 and 4 show that the strategies differed primarily in the direction of their scoring errors: FS+CoT tended to under-score, RAG tended to over-score, and SC showed little mean directional bias despite substantial response-level discrepancies.

\begin{figure}[htbp]

\centering

\includegraphics[width=0.95\textwidth]{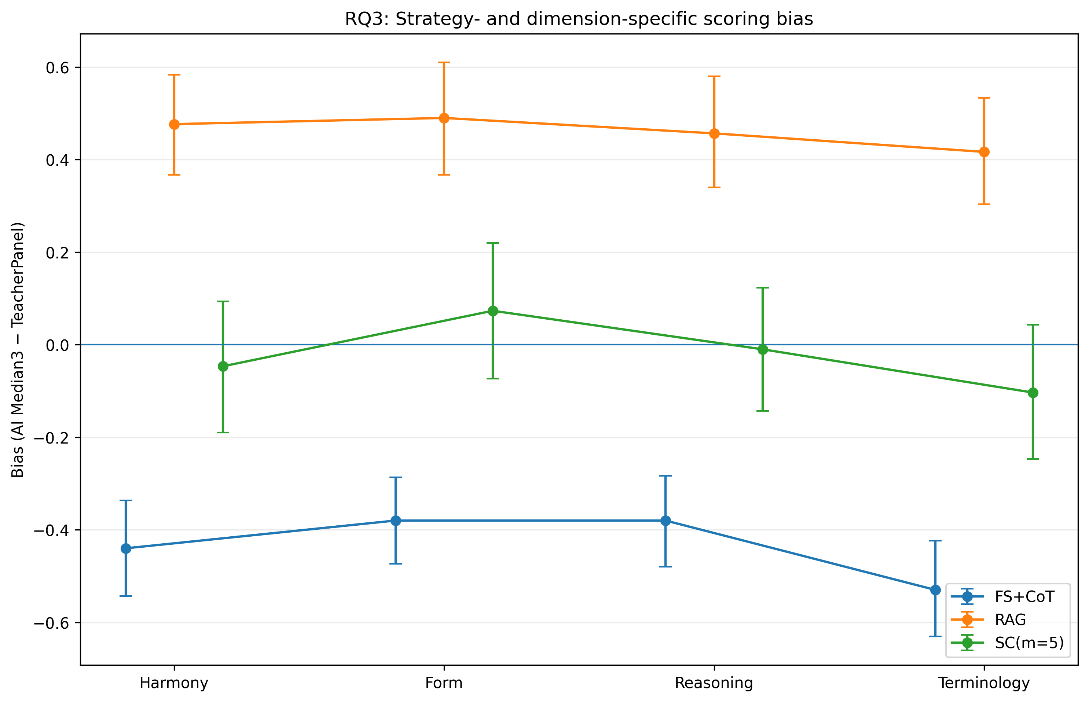}

\caption{Strategy and dimension scoring bias}

\end{figure}

\begin{figure}[htbp]

\centering

\includegraphics[width=0.95\textwidth]{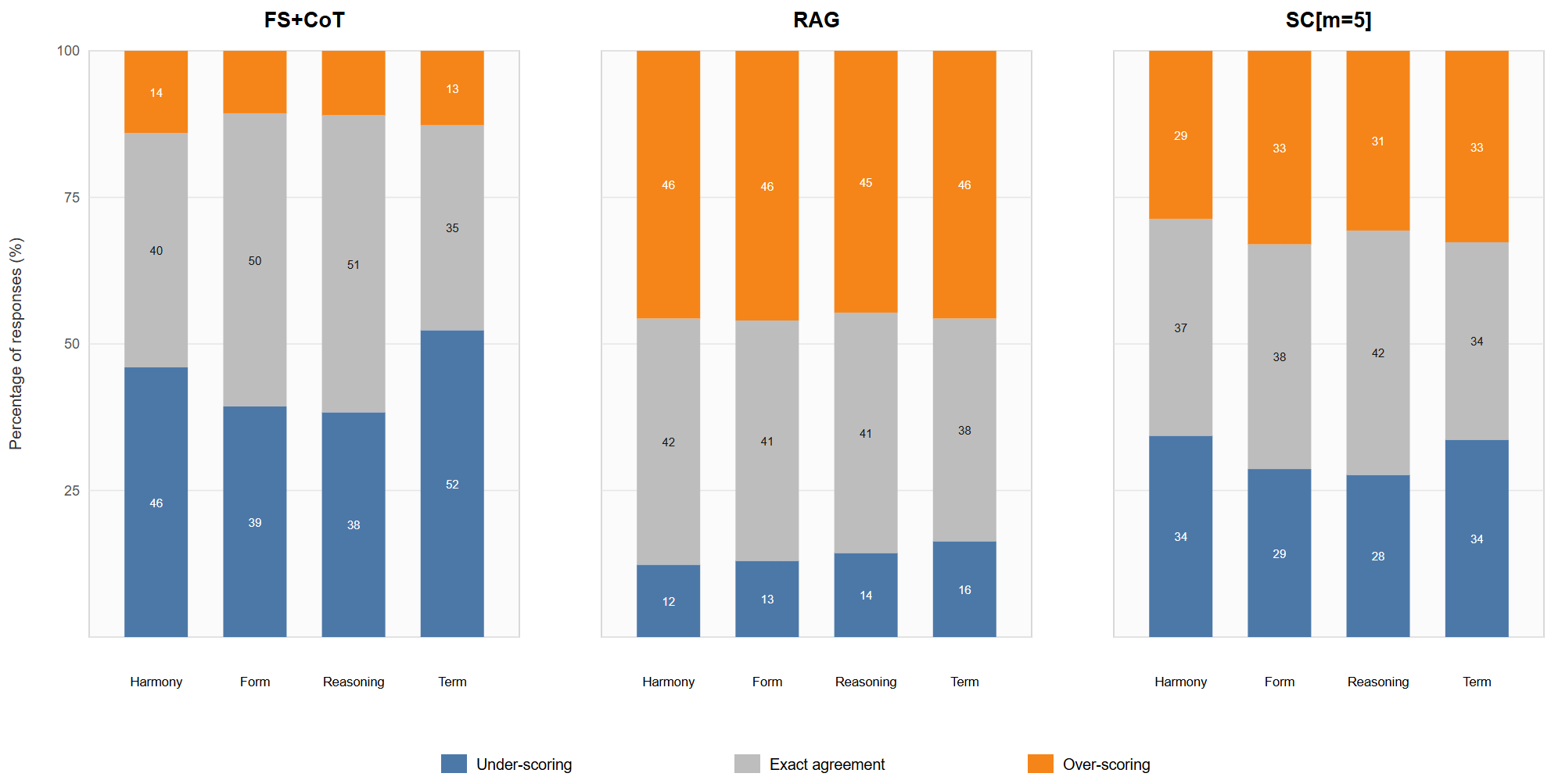}

\caption{Direction of scoring differences relative to teacher mean scores}

\end{figure}

\section{Discussion and Implications}

The three prompting strategies produced distinct patterns of agreement, error, and systematic bias. Among the three strategies examined, Fs+CoT showed the strongest alignment with the teacher mean scores, RAG exhibited systematic leniency, and SC produced highly repeatable but less accurate individual-level scores.

\subsection{Prompting Strategies as Distinct Scoring Behaviors}

The comparatively strong performance of Fs+CoT suggests that scored examples and structured reasoning can help an LLM approximate the analytic judgments of human teachers. Few-shot examples may have served as scoring anchors, while the reasoning instructions encouraged the model to examine multiple aspects of the rubric before assigning a score. This combination appears especially useful for dimensions requiring the integration of textual evidence and evaluative judgment. RAG displayed the opposite pattern. The retrieval of disciplinary information may have increased the model's sensitivity to the presence of relevant musical concepts and terminology. However, the model may have awarded credit when students mentioned relevant concepts without applying them accurately or explaining them sufficiently. Retrieval can therefore strengthen access to domain knowledge without guaranteeing that the model will apply the same quality standards as expert teachers. For scoring purposes, additional knowledge is beneficial only when it is accompanied by appropriate evaluative thresholds. SC was distinctive in showing little overall directional bias while retaining relatively large individual-level errors. The difference between group-level calibration and response-level agreement. Over-scoring some responses and under-scoring others can produce an average bias close to zero, even when many individual scores differ from teacher judgments. SC was therefore balanced in aggregate score location but not consistently accurate for individual students.

\subsection{Repeatability Does Not Equal Agreement}

All three strategies showed relatively stable scoring patterns across repeated runs. Similar observations have been reported in recent LLM evaluation studies, where aggregating multiple independent model outputs primarily reduced sampling variability while rarely changing the substantive ranking of prompting strategies \citep{wang2022selfconsistency,manakul2023selfcheckgpt}. SC provides the clearest example of why repeatability and agreement must be evaluated separately. It generated highly similar scores across runs, yet its agreement with teacher mean scores remained weaker than that of Fs+CoT. A scoring system can consistently reproduce its own decisions while consistently differing from expert judgments. Repeatability is necessary for dependable scoring, but it does not establish agreement with expert judgments or support score interpretation on its own. Median aggregation produced only limited improvement because Run1 scores were already relatively stable. More importantly, aggregation did not correct the systematic severity of Fs+CoT or the leniency of RAG. Aggregation can reduce random variation and the influence of occasional extreme outputs, but it cannot correct a directional bias that recurs across runs.

Repeated scoring increases computational cost and processing time, but it may provide little additional benefit when a strategy is already stable. The use of multiple runs should therefore be justified empirically for each prompting condition rather than assumed to improve scoring automatically. Calibration and prompt refinement may be more important than simply increasing the number of generations.

\subsection{Dimension-Specific Scoring Challenges}

The variation across rubric dimensions indicates that LLM scoring performance depends on the nature of the construct being evaluated. When students connected evidence to conclusions, explained relationships among musical elements, or justified their interpretations, the prompting strategies captured aspects of their analytical reasoning more consistently than their use of disciplinary terminology. Terminology was more difficult to evaluate consistently. Although technical terms can be identified lexically, accurate terminology use requires more than word recognition. Students may name the correct concept but apply it incorrectly, use it without sufficient explanation, or confuse related musical functions. Human teachers may be more sensitive than the model to the distinction between mentioning a technical term and applying it accurately.

The importance of dimension-level validation. Total-score agreement may conceal weaknesses in particular dimensions because strong performance in one dimension can compensate numerically for poor performance in another. Analytic automated-scoring studies have therefore emphasized the need to evaluate each rubric dimension separately, particularly when dimensions involve different linguistic, conceptual, and evidential demands \citep{rahimi2017assessing,yoo2025dataset}.

\subsection{Implications for Music Analysis Assessment}

For music education, the results support the use of LLM scoring as a supplementary rather than autonomous assessment tool. A practical implementation would combine automated scoring with human oversight. The LLM could provide preliminary scores or criterion-referenced feedback and flag responses that require further review. Teachers could then examine cases involving unusual score patterns, uncertain terminology, large discrepancies among dimensions, or potential over- and under-scoring. Such a human-in-the-loop model would preserve the efficiency of automated scoring while retaining expert judgment for decisions requiring expert interpretation.

The findings are particularly relevant for written music-analysis tasks because these assessments require students not only to identify musical features, but also to explain relationships, justify interpretations, and use disciplinary terminology accurately. LLMs may support the evaluation of these elements, but they do not yet apply all rubric dimensions with equivalent precision.

\subsection{Overall Implications and Limitations}

LLM-based scoring should be evaluated not only by agreement coefficients, but also by systematic bias, dimension-specific performance, and cross-run repeatability. Among the strategies examined, Fs+CoT was the most promising for scoring written music analyses; however, calibration and human oversight remain necessary before operational use.

This study examined only GPT-4o-mini on a single music-analysis task from one institutional context, using three runs at a temperature of 0.9. Although the teacher mean scores represented the students' final grades and served as the formal reference for evaluation, the findings may not generalize to other models, tasks, sampling settings, institutions, or high-stakes assessment contexts.

\end{document}